\documentclass{Rinton-P9x6}

\begin{document}

\title{Mass scales in the RS models: the $\mu$ and cosmological 
constant\footnote{Talk presented at CICHEP, Cairo, Egypt,
Jan. 9--13, 2001.}}

\author{Jihn E. Kim}

\address{Dept. of Physics and Center for Theoretical Physics,
Seoul National University, Seoul 151-747, Korea
\\E-mail: jekim@phyp.snu.ac.kr}


\maketitle

\abstracts{There is the scale problem in the Randall-Sundrum 
models. Regarding this, I review the works done with B. S. Kyae and
H. M. Lee. In a supersymmetric generalization in the RSI model, 
we discuss that the $\mu$ parameter
can be obtained by an intermediate scale brane.\cite{kyae} 
In addition, we also discuss the cosmological
constant problem with a self-tuning solution
in the RSII model.\cite{kkl1,kkl2}}

\def\p{\partial}

The standard model has been extremely successful phenomenologically.
However, it has 20 theoretically unexplained parameters, among which
the Higgs scalar mass is called the gauge hierarchy 
problem: Why is the electroweak scale $10^{-16}$ times the
Planck mass? To solve this hierarchy problem, technicolor and
supersymmetry have been extensively studied in the last twenty years.

Recently, there appeared another try toward understanding this
gauge hierarchy problem, through large extra dimensions or through
warp factor geometry in the so-called RSI model~\cite{rs1}. There
are two branes located at $y=0$(B1 brane) and $\pi$(B2 brane)  
where $y$ is the fifth coordinate. The B1 brane tension 
$\Lambda_1=6k_1M^3>0$, B2 brane tension $\Lambda_2=6k_2M^3<0$ and 
the bulk cosmological constant $\Lambda_b=-6k^2M^3<0$ are fine-tuned
$k_1=k=-k_2$ to have a flat geometry, where $M$ is the fundamental
mass parameter in 5D. With this kind of two fine-tunings
the flat-space solution is possible even if $R^2$ terms are added
if their form is of the Gauss-Bonnet type~\cite{kkl3}. 

The RSI model imposes the symmetry $S_1/Z_2$ for compactification,
with the flat space metric ansatz, $ds^2=e^{-2\sigma(y)}
dx_\mu dx^\mu + r_c^2dy^2$, which allows the solution
$\sigma=kr_c|y|$. Integrating over $y$, we obtain an effective 
4D Planck mass, $M_P=(M^3/k)(1-e^{-kr_c})$ which is again of
order $M$. This fundamental mass parameter governs the mass scale
at $y=0$. On the other hand, at B2($y\ne 0$) the rescaling of the fields so
that the standard kinetic energy terms result gives the mass parameter
of order $m=M_{input}e^{-kr_c/2}$ which can be interpreted as a
TeV scale if $r_c$ is a few tens of $M^{-1}$. Thus, interpreting
B2 as the brane for housing the visible sector fields, we obtain
a long anticipated exponetially suppressed electroweak mass scale
compared to the Planck mass. This led to a stimulus since it can
be another
solution for the gauge hierarchy problem. Note that here the key
point is that a warp factor is introduced, i.e. a cuved space in
$y$ direction even if 4D is flat.

However, this interesting gauge hierarchy solution has a
different kind of scale problem. At B2 the ultimate mass scale
is TeV. Thus, the gauge coupling unification at GUT scale is
not possible, which was shown to be so remarkable in
$SO(10)$ by Raby~\cite{raby}. Unification may be achieved above
TeV scale by host of KK modes~\cite{dienes} (which seems to be a 
fitting rather than prediction) but then the proton decay
operator has the relevant mass parameter of order TeV which
makes proton lifetime tens of orders shorter than the present
bound. In addition, the TeV scale does not introduce the needed 
very light axion for the strong CP solution~\cite{axion}. The mm scale
gravity can introduce a very light axion, but it needs a several
internal dimensions~\cite{chang}, not achievable in 5D RSI model. 
Also, the transition from inflationary phase to the Big Bang
phase needs the visible sector brane tension positive~\cite{cline}, 
which is not the case in RSI model. Thus, it seems that RSI model,
designed for the gauge hierarchy solution, introduces another
kind of the mass scale problem.

In this talk, I present two possible solutions for the mass scale
in the Randall-Sundrum models. One is the $\mu$ parameter which we
try to understand in the RSI model, and the other is the 
cosmological constant which we try to understand in the RSII
model~\cite{rs2}.

\vskip 0.5cm
\centerline{The $\mu$ in RSI model}
\vskip 0.1cm

Because of the scale problems we encounter in the RSI model, we
do not introduce a TeV scale brane. Then the original beautiful
motivation for the gauge hierarchy is lost. So, the gauge hierarchy
is understood by the conventional supergravity theory. Then, there
exists another scale problem: the $\mu$ problem~\cite{kn}.
In this case, we put visible sector fields at the Planck brane(B1).
The second brane(B2) is interpreted as an intermediate scale
brane. There is no serious cosmological problem since the brane
tension at B1 is positive. A logarithmic unification of gauge 
couplings is possible because the visible brane is the Planck
scale brane. Since the KK modes are massive at the intermediate
scale, the low energy physics is very similar to the MSSM.
Strong CP solution is possible by introducing a very light 
axion. 

 We introduce the chiral fields at B1, bulk and B2 as shown in
Table 1, where $A$ is the global axial charge, corresponding to
the Peccei-Quinn charge. We have not specified the graviton
multiplet in the bulk. The MSSM gauge multiplets live at B1.

\begin{table}[t]
\caption{The B1, bulk and B2 fields and their global
charges\label{tab:charge}}
\begin{center}
\footnotesize
\begin{tabular}{|c|c|c|l|}
\hline
{Brane or bulk} &\raisebox{0pt}[13pt][7pt]B1 &
\raisebox{0pt}[13pt][7pt]Bulk &{B2}\\
\hline
\multicolumn{1}{|c|}{\raisebox{0pt}[12pt][6pt]{Fields}} 
& $\ H_1\ \ H_2\ \ MSSM$ &$\Sigma^i(i=1,2)$ &$\ S\ \ Z$\\
\cline{1-4}
$A$   &$-1\  -1\ \ \ \ +\frac{1}{2}\ \ $ 
&\begin{minipage}{1in}
\begin{center}
$+1$ 
\end{center}\end{minipage} 
&\begin{minipage}{1.5in}
\phantom{xxx}
$-1\ \ 0$\\
These fields located at $y\ne 0$ brane are required
to couple supersymmetrically to $\Sigma^i$.
\end{minipage} \\[22pt]
\hline
\end{tabular}
\end{center}
\end{table}

If N=1 supersymmetry is present at B1, we can write a
superpotential at B1 as
\begin{equation}
W\sim \frac{\Sigma^2}{M_P}H_1H_2,\label{nonren}
\end{equation} 
which is a schematic formula. The exact meaning will be
given after what are the zero mass chiral fields at
low energy.
If bulk and brane fields respect the N=1 supersymmetry, then at
B2 we anticipate a superpotential
\begin{equation}
W^\prime\sim Z(\Sigma S-M_P^2),
\end{equation}
which is again a schematic formula.
But because B2 is the intermediate scale brane, the VEV of $\Sigma$
would be an intermediate scale, and in view of Eq.~(\ref{nonren})
we expect a TeV scale $\mu$.

The 5D bulk action is
\begin{eqnarray}
&S_{bulk}=-\Sigma_i\int d^5 x\sqrt{-G}\Large[g^{MN}\p_M
\Phi^{i *}\p_N\Phi^i+(\bar\Psi\gamma^M\nabla_M\Psi\nonumber\\
&-(\nabla_M\bar\Psi)\gamma^M\Psi)+M^2_{\Phi^i}|\Phi^i|^2
+M_\Psi\bar\Psi_L\Psi_R+M_\Psi\bar\Psi_R\Psi_L
\Large]
\end{eqnarray}
where $\gamma^M\equiv e^M_a\gamma^a,\nabla_M=\p_M+\Gamma_M,
\Gamma_\mu=\frac{1}{2}\gamma_5\gamma_\mu\frac{d\sigma}{dy},
\Gamma_5=0$, which satisfies the N=2 supersymmetry~\cite{gp}.
In the $AdS_5$ the fields in the same hypermultiplet must
have different masses with one undetermined parameter
$t$,
\begin{eqnarray}
&M^2_{\Phi^1}=(t^2+t-\frac{15}{4})\sigma^{\prime 2}+
(\frac{3}{2}-t)\sigma^{\prime\prime}\nonumber\\
&M^2_{\Phi^2}=(t^2-t-\frac{15}{4})\sigma^{\prime 2}
+(\frac{3}{2}+t)\sigma^{\prime\prime}\\
&M_\Psi=t\sigma^\prime\nonumber
\end{eqnarray}
where it is obvious that the fermion mass is odd under the
$Z_2$ parity and the boson masses are even under the $Z_2$
parity.

The 5D field equation for the hypermultiplet is
\begin{equation}
\left[e^{2\sigma}\eta^{\mu\nu}\p_\mu\p_\nu+
e^{s\sigma}\frac{\p}{\p y}e^{-s\sigma}\frac{\p}{\p y}
-M^2_\Phi\right]\Phi(x^\mu,y)=0
\end{equation}
where $s=4$ for a scalar $\phi$ and $s=1$ for the fermion 
$\psi_{L,R}$. Thus, $M^2_\Phi=ak^2+b\sigma^{\prime\prime}$
and $M^2_\Phi=k^2t(t\pm 1)\mp t\sigma^{\prime\prime}$ for $\Psi_{L,R}$.
The KK mode decomposition
\begin{equation}
\Phi(x^\mu,y)=\frac{1}{\sqrt{2b_0y_c}}
\sum_{n=0}^\infty\Phi^{(n)}(x^\mu)f_n(y)
\end{equation}
gives a solution for $f_n(y)$ and hence the KK masses. For 
$n\ge 1$, we obtain massive KK modes. The odd fields do have
massless modes due to the orbifold condition. It is easy to check
that massless condition $\hat M^2=e^{s\sigma}\frac{\p}{\p y}
e^{-s\sigma}\frac{\p}{\p y}$ gives the following pair of
the massless modes from $Z_2$ even field,
\begin{equation}
\Phi^{1,(0)}(x,y)=\frac{e^{(\frac{3}{2}-t)\sigma(y)}}{\sqrt{2b_0y_c}N}
\phi(x)_\Sigma,\ \ \Psi_L^{(0)}(x,y)=
\frac{e^{(2-t)\sigma(y)}}{\sqrt{2b_0y_c}N}
\psi_\Sigma(x)
\end{equation}
where $N^2=(e^{\sigma_c(1-2t)}-1)/\sigma_c(1-2t)$ with $\sigma_c=kb_0
y_c$. Note that $N(t=\frac{1}{2})=1$. One can explicitly show that
the above $n=0$ KK modes are massless for any $t$. Let us call them
$\phi_\Sigma,\psi_\Sigma$, respectively. These massless bulk fields
couple to the B2 brane fields $S$ and $Z$. Here comes the fixing of
$t$. Only for $t=1/2$ the couplings at B2 are maintained to be
supersymmetric~\cite{kyae}.

Thus, we just show the result for $t=1/2$. The B2 fields with a
proper normalization are called tilde fields. Thus, the massless
bulk $\Sigma$ field becomes $\tilde\Sigma$ at B2
\begin{equation}
\bar\Sigma=\{e^{\sigma(y)}\tilde\phi_\Sigma(x),e^{\frac{3}{2}\sigma(y)}
\tilde\psi_\Sigma(x)\}\rightarrow \tilde\Sigma(x)
=\{\tilde\phi_\Sigma(x),\tilde\psi_\Sigma(x)\}.
\end{equation} 

At B2, $\phi_i\sim e^{\sigma_c}\tilde\phi_i$ and $\psi_i\sim
e^{3\sigma_c/2}\tilde\psi$ couple to brane fields $S=\{\phi_S,
\psi_S\}$ and $Z=\{\phi_Z,\psi_Z\}$. For example, the following
couplings result,
\begin{eqnarray}
&S^{int}_{B2}=\int d^4x\sqrt{-g_4}\Big[
\{(e^{\sigma_c}\tilde\phi_\Sigma)\psi_S\psi_Z+
\phi_S\psi_Z(e^{3\sigma_c/2}\tilde\psi_\Sigma)+\phi_Z(e^{3\sigma_c/2}
\tilde\psi_\Sigma)\psi_S\nonumber\\
&+h.c.\}-|\phi_S\phi_Z|^2-|\phi_Z(e^{\sigma_c}\tilde\phi_\Sigma)|^2
-|(e^{\sigma_c}\tilde\phi_\Sigma)\phi_S|^2
+M_P^2((e^{\sigma_c}\tilde\phi_\Sigma)\phi_S\nonumber\\
&+h.c.)-M_P^4 \Big]\ =\ 
\int
d^4x\Big[(\tilde\phi_\Sigma\tilde\psi_S\tilde\psi_Z+\tilde\phi_S
\tilde\psi_Z\tilde\psi_\Sigma+\tilde\phi_Z\tilde\psi_\Sigma\tilde\psi_S
\nonumber\\
&+h.c.)-|\tilde\phi_S\tilde\phi_Z|^2-|\tilde\phi_Z\tilde\phi_\Sigma|^2
-|\tilde\phi_\Sigma\tilde\phi_S-M_I^2|^2\Big]\label{susy}
\end{eqnarray}
where $M_I=M_Pe^{-\sigma_c}\sim 10^{11-13}$~GeV for $\sigma_c\simeq
11.5-16$. This interaction (\ref{susy}) is obtained from a
superpotential $W=\tilde Z(\tilde\Sigma\tilde S-M_I^2)$. This kind
of supersymmetric interaction is possible for $t=1/2$. $W$ guarantees
the VEV of $\tilde\Sigma$ and $\tilde S$ is of order $M_I$ when we
consider the soft terms generated by supergravity. Therefore, the
Peccei-Quinn symmetry is broken at the intermediate
scale and there results a very light axion. At B1, the $\mu$ term
is generated at the order
\begin{equation}
\mu=\frac{\tilde\Sigma^2}{M_P}\sim m_{3/2}.
\end{equation}

We have an $N=1$ supersymmetric theory, with the $\mu$ term 
generated at B2 via the PQ symmetry breaking at the intermediate
scale $M_I$. Since the SUSY breaking
scale in SUGRA is similar to the PQ symmetry breaking scale, it
is desirable to break supersymmetry at B2 since the natural
scale at B2 is $M_I$. It can be achieved by the gaugino
condensation at B2, presumably $E_8^\prime$ gaugino condensation.
With the fundamental scale $M$, the condensation scale is very
close to $M$ since the $\beta$ function of $E_8^\prime$ is large and
negative. Thus, at B2 the condensation scale of $E_8^\prime$ is
of order $M_I$, and hence breaking supersymmetry at order $M_I$. Thus,
the gravitino mass $m_{3/2}$ is of order electroweak scale.
One can imagine that the soft mass generation of a sfermion 
at B1 proceeds through its coupling 
to gravitinos at B1, the gravitino propagation 
in the bulk and the gravitino mass generation at B2. However, it
is more reliable to integrate an effective action with respect to $y$
to obtain the soft mass~\cite{kyae}.
In this way, one obtains the soft masses of the MSSM fields at B1 
at order TeV. 

A common supersymmetry breaking scale and the axion scale is an
intreaguing hypothesis, proposed several times with different
contexts~\cite{common,kn}. Here, we achieved it by introducing
an intermediate scale brane B2.

\vskip 0.5cm
\centerline{\bf Self-tuning solution of the cosmological 
constant problem}
\vskip 0.1cm

The cosmological constant problem can be addressed from early
1920's. But it became the modern particle physics problem 
since the spontaneous symmetry breaking in particle physics
was accepted as the mass generation mechanism~\cite{veltman}.
The VEV of potential or $-{\cal L}$ is interpreted as the
cosmological constant, $\langle-{\cal L}\rangle=V_0
\equiv\Lambda_{eff}$.

The flat space solution is possible with $V_0=0$, while de
Sitter space solution and anti de Sitter space solution 
result with $V_0>0$ and $V_0<0$, respectively. By measuring
the curvature of the universe~\cite{perl}, it is known that $|V_0|
\le (0.003\ {\rm eV})^4$ which needs an extreme fine tuning
of parameters in the Lagrangian which are supposed to be
described by parameters at the fundamental mass scale,
i.e. at the Planck mass scale $M_P$. $\lq\lq$Why is $V_0$ so 
accurately fine tuned?", which is the cosmological constant
problem. This problem is very severe in spontaneous symmetry
breaking models which introduce vacuum energies in the process
of seeking the true minimum of the potential.

Toward the solution of the problem, self-tuning idea has been
introduced by Witten~\cite{witten} and Hawking~\cite{hawking}.
Their definition is $\lq\lq$the existence of the flat space solution
without any fine-tuning of parameters for a finite range of 
parameter space in the Lagrangian." It was
different from the current usage of self tuning~\cite{kachru}, 
in which they need only a flat space solution excluding the
possibility of de Sitter and anti de Sitter space solutions.
Here, we adopt the earlier definition\cite{witten,hawking} 
since in this case one does not rule out the possibility of 
inflation. Furthermore, one has to resolve a small cosmological
constant reported recently~\cite{perl}, for which quintessence
ideas were proposed~\cite{quint}.

In the RSI model a flat space solution is possible with two fine
tunings, $k_1=k=-k_2$, where $k^2=-\Lambda_b/6, k_1=\Lambda_1/6$
and $k_2=\Lambda_2/6$ in terms of the bulk cosmological
constant $\Lambda_b$, tension $\Lambda_1$ at B1, and tension 
$\Lambda_2$ at B2. In the second Randall-Sundrum 
model(RSII)~\cite{rs2}, there is only one brane located at
$y=0$, which is called B1. Here, the flat space is obtained
by a fine-tuning $k_1=k$. Note, however, that these models
start with nonvanishing $\Lambda$'s but allow flat space
solutions, which gives a hope for obtaining a model for
vanishing cosmological constant.

In the recent attemts, the study is limited to a classical
action only~\cite{kachru}. The bulk potential is coupled to
a function $f(\phi)$ of a scalar field $\phi$, satisfying the
condition $(d/d\phi)f(\phi)=f(\phi)$, which can be thought of
a fine-tuning or not, depending on one's judgement. In this case,
it has been known that one fine-tuning is needed, as explained
below. 
In general, the
self-tuning solutions need to satisfy: (i) There should exists an
undetermined integration constant so that it self-tunes the
cosmological constant, and (ii) there should not appear a
naked singularity within the allowed region of space-time.
The new attempt will require in addition that there should not
exist de Sitter and anti de Sitter space solutions. Even though
the recent attempt~\cite{kachru} obtains an undetermined
integration constant, it has a naked singularity. Resolving the
naked singularity by putting a brane there requires a fine tuning.  
It is easy to understant this fine-tuning. As soon as there is a
need to insert a brane, there appears the brane tension as
a free parameter. This parameter must be fine-tuned to give
a flat space since the integration with respect to $y$ pick up
this brane tension as an additional vacuum energy. If total
effective vacuum energy were zero at one value of the brane
tension, the total vacuum energy would not be zero for another 
value of the brane tension. Therefore, the need to introduce
another brane necessarily returns to a fine-tuning case.

Here we report a recent work~\cite{kkl1,kkl2} which allows 
self-tuning of the cosmological constant. 
We work in the RSII type set up, i.e. one 3-brane is located at
$y=0$ in 5D. Matter fields reside at the brane. In the bulk
we introduce the graviton and a three
index antisymmetric tensor field $A_{MNP}$ whose field strength is
$H_{MNPQ}$. The standard $H^2$ term does not give a self-tuning
solution. Thus, we are led to consider the $1/H^2$ term, or
more generally $1/(H^2)^n\ (n\ge 1)$ terms. 
The ansatz for the flat space is $ds^2=\beta(y)\eta_{\mu\nu}
dx^\mu dx^\nu+dy^2$. For the case of $1/H^2$ term, the field 
equations are satisfied with the flat space solution~\cite{kkl1},
\begin{equation}
\beta(|y|)=\left(\frac{k}{a}\right)^{1/4}
\left(\frac{1}{\cosh(4k|y|+c)}\right)^{1/4}
\end{equation} 
where the undetermined integration constant is 
\begin{equation}
c=\tanh^{-1} \frac{\Lambda_1}{\sqrt{-6\Lambda_b}}
\end{equation}
where $\Lambda_1$ is the brane tension of the brane located
at $y=0$ and $\Lambda_b$ is the bulk cosmological constant.
Its shape is shown in Fig. 1.
It is easy to see that for a finite range of the brane 
tension an integration constant can be determined to give
the above flat space solution. Thus, we realize Witten and
Hawking's self-tuning idea. Note that there also exist
the anti de Sitter space and de Sitter space solutions,
which are shown in Figs. 2 and 3. But Hawking's most
probable universe chooses the vanishing cosmological
constant, even if the electroweak or QCD phase transitions
add a constant to an initial $\Lambda_1$~\cite{kkl2}.

\begin{figure}[t]
\epsfxsize=15pc
\epsfbox{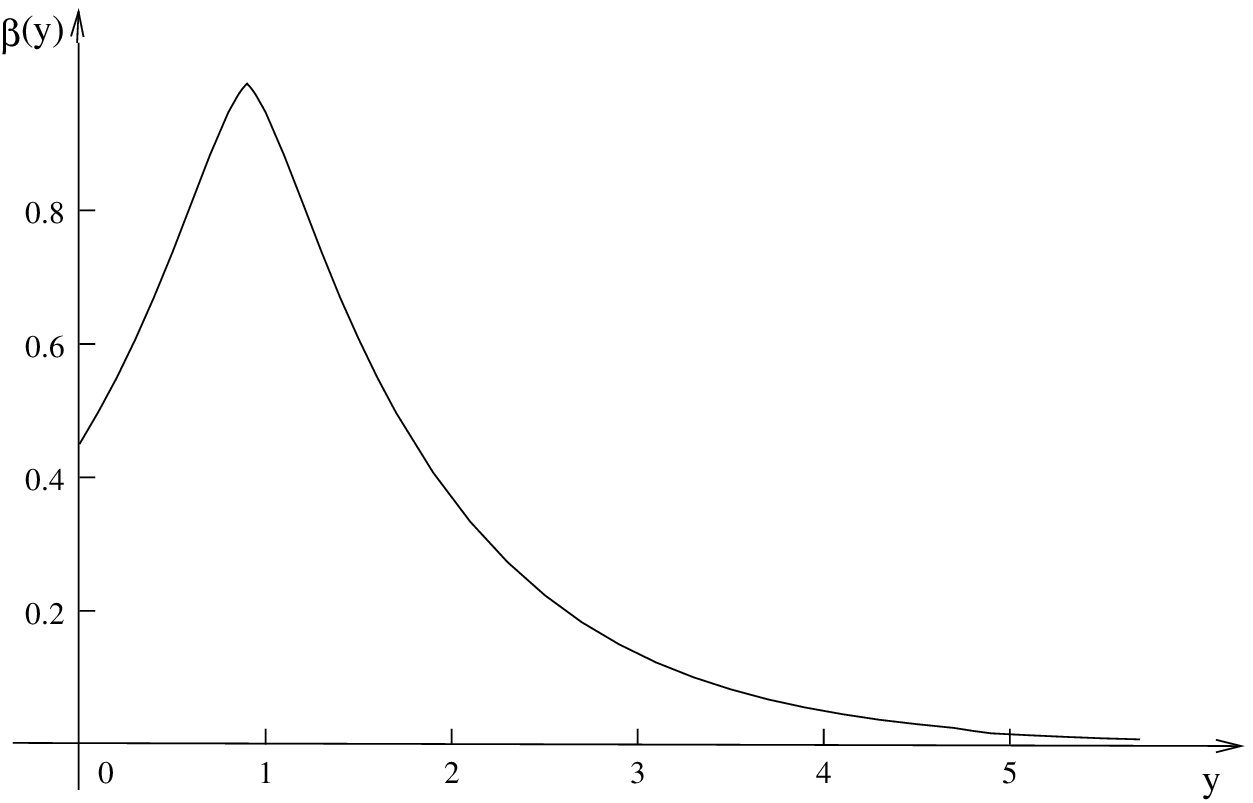} 
\caption{The flat space solution.
  \label{fig:beta0}}
\end{figure}
\begin{figure}[t]
\epsfxsize=15pc
\epsfbox{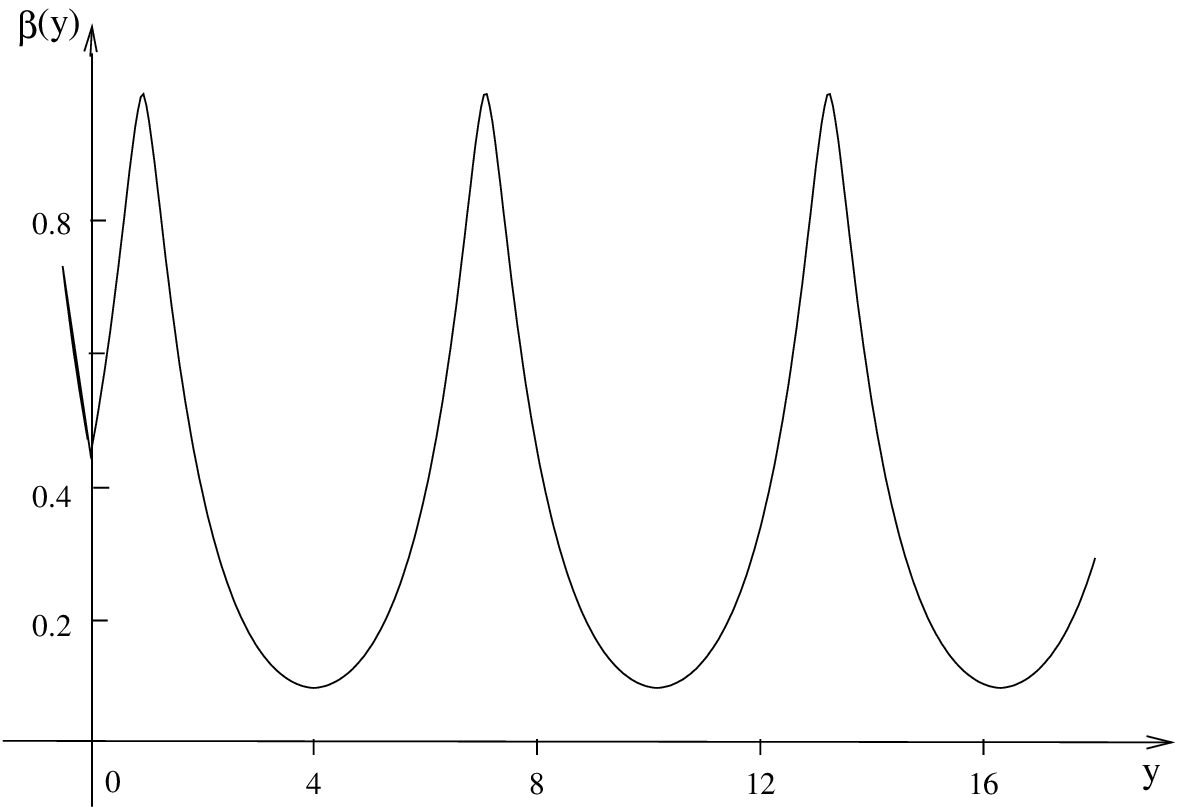} 
\caption{The anti de Sitter space solution.
  \label{fig:beta-}}
\end{figure}
\begin{figure}[t]
\epsfxsize=15pc
\epsfbox{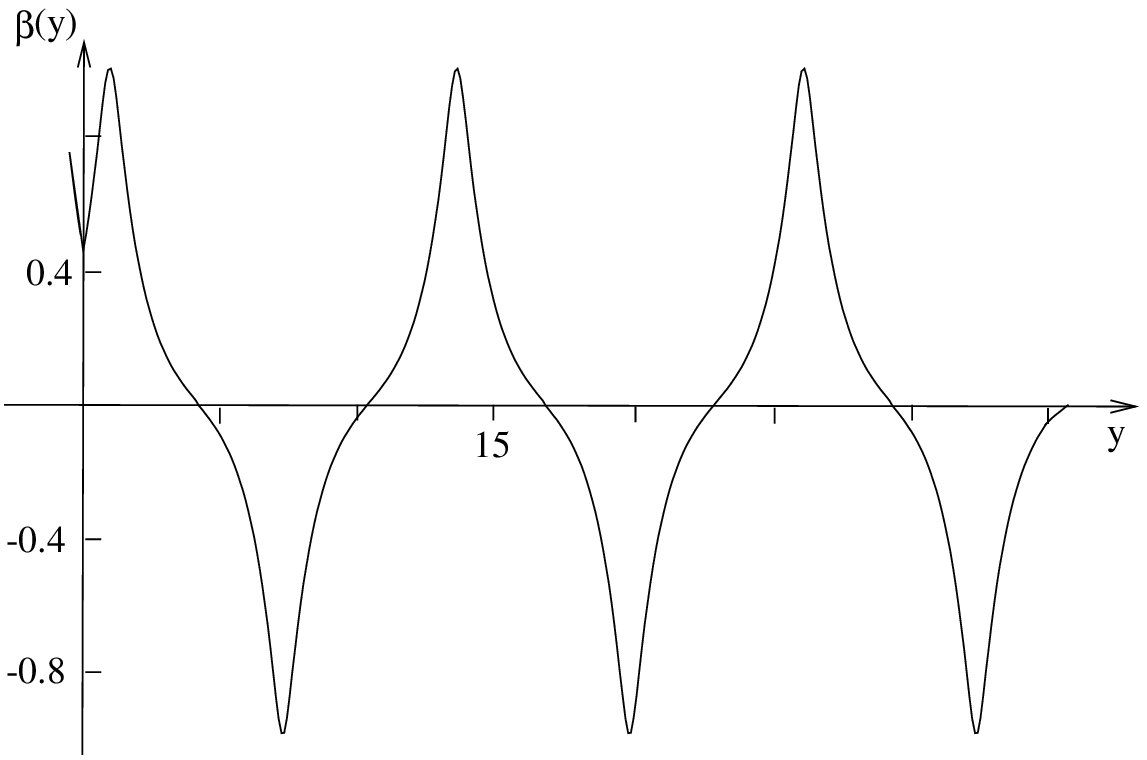} 
\caption{The de Sitter space solution.
  \label{fig:beta+}}
\end{figure}

We have seen that the Randall-Sundrum models offer possibilities
for solving some mass hierarchy problems, in particular the
$\mu$ problem in a RSI type model and the cosmological constant
problem in a RSII type model.

\section*{Acknowledgments}
This work is supported in part by the BK21 program of Ministry 
of Education, Korea Research Foundation Grant No. KRF-2000-015-DP0072, 
and by the Center for High Energy Physics(CHEP),
Kyungpook National University.

\end{document}